\documentclass[conference,10pt]{IEEEtran}
\usepackage{cite}
\usepackage{amsmath,amssymb,amsfonts}
\usepackage{algorithmic}
\usepackage{graphicx}
\graphicspath{ {figs/} }
\usepackage{siunitx}
\usepackage{xcolor}
\usepackage{xspace}
\usepackage{todonotes}
\usepackage{balance}
\usepackage{listings}

\usepackage{pxfonts}
\usepackage[T1]{fontenc}

\ifCLASSOPTIONcompsoc
    \usepackage[caption=false,font=normalsize,labelfont=sf,textfont=sf]{subfig}
\else
    \usepackage[caption=false,font=footnotesize]{subfig}
\fi

\newcommand{\mynote}[3]{%
}

\newcommand\djb[1]{\mynote{Dan}{0,0,255}{ #1}}
\newcommand\fz[1]{\mynote{Foivos}{81,13,74}{ #1}}
\newcommand\vk[1]{\mynote{vk}{0,100,0}{ #1}}

\newcommand{\numascope}{{NUMAscope}\xspace}
\newcommand{\opensource}{open-source\xspace}

\begin{document}
\title{\numascope: Capturing and Visualizing Hardware Metrics on Large ccNUMA Systems}

\author{
     \IEEEauthorblockN{Daniel J Blueman}
     \IEEEauthorblockA{\textit{Principal Software Engineer} \\
     \textit{Numascale AS}\\
     Oslo, Norway\\
     daniel@numascale.com}
 \and
     \IEEEauthorblockN{Foivos Zakkak}
     \IEEEauthorblockA{\textit{Department of Computer Science} \\
     \textit{The University of Manchester}\\
     Manchester, United Kingdom\\
     foivos.zakkak@manchester.ac.uk}
 \and
     \IEEEauthorblockN{Christos Kotselidis}
     \IEEEauthorblockA{\textit{Department of Computer Science} \\
     \textit{The University of Manchester}\\
     Manchester, United Kingdom\\
     christos.kotselidis@manchester.ac.uk}
}

\maketitle
\thispagestyle{plain}
\pagestyle{plain}

\begin{abstract}
    Cache-coherent non-uniform memory access (ccNUMA) systems enable parallel applications to scale-up to thousands of cores and many terabytes of main memory.
    However, since remote accesses come at an increased cost, extra measures are necessitated to scale the applications to high core-counts and process far greater amounts of data than a typical server can hold.
    In a similar manner to how applications are optimized to improve cache utilization, applications also need to be optimized to improve data-locality on ccNUMA systems to use larger topologies effectively.
    The first step to optimizing an application is to understand what slows it down.
    Consequently, profiling tools, or manual instrumentation, are necessary to achieve this.
    When optimizing applications on large ccNUMA systems, however, there are limited mechanisms to capture and present actionable telemetry.
    This is partially driven by the proprietary nature of such interconnects, but also by the lack of development of a common and accessible (read \opensource) framework that developers or vendors can leverage.
    
    In this paper, we present an \opensource, extensible framework that captures high-rate on-chip events with low overhead (\textless 10\% single-core utilization).
    The presented framework can operate in \texttt{live} or \texttt{record} mode, allowing both \textit{real-time} monitoring or capture for later post-workload or offline analysis.
    High-resolution visualization is available either through a \textit{standards-based} (web) interactive graphical interface or through a convenient textual interface for \textit{quick-look} analysis. 
\end{abstract}



\section{Introduction}
Cache-coherent non-uniform memory access (ccNUMA) architectures enable applications to scale up to hundreds of cores and terabytes of memory without the need of explicit synchronization or communication between the different boards of the system.
ccNUMA systems essentially provide a single system image (SSI) to the software, abstracting away the memory hierarchy and thus easing development.
To achieve this, ccNUMA systems rely on cache-coherent interconnects and memory-coherence protocols.
This approach however may come at a cost if applications do not account for the overheads of remote accesses on a ccNUMA system.

\begin{figure}
    \includegraphics[clip,width=\linewidth]{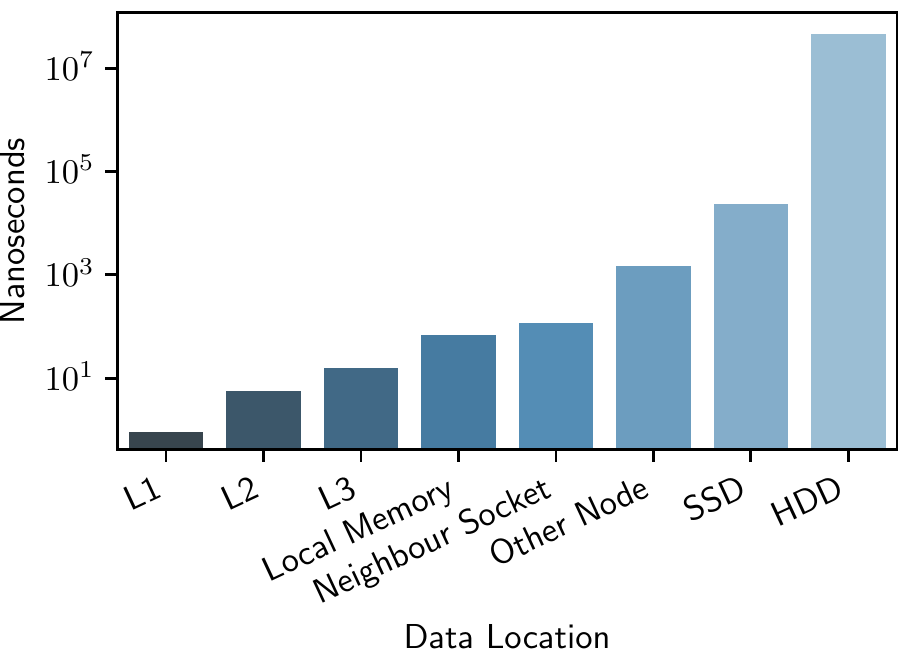}
    \caption{Latency of memory accesses depending on the distance between the core performing the access and the memory being accessed}
    \label{fig:numa-costs}
\end{figure}

Although memory-coherence is implemented in hardware and is transparent to the software, it comes with overhead which need to be taken into account when optimizing an application; the memory hierarchy is such that main memory access time differs depending on the memory segment being accessed and the core that performs the access.
Figure~\ref{fig:numa-costs} demonstrates how distant memory accesses relate to the latencies of accessing different levels of the memory hierarchy.
As a result in a similar way that parallel applications should avoid cache line sharing (so as to avoid excessive coherence traffic), they should minimize writing cache lines frequently read among many NUMA nodes~\cite{hackenberg2009comparing}.

The first step in avoiding high latency accesses is to quantify them by inspecting the system's performance counters.
However, when developing and tuning software applications on systems large enough to warrant cache-coherent interconnects, existing mechanisms to collect resource usage are limited to the system's core and uncore performance counters.
For instance, existing NUMA-aware profilers can attribute increased wait cycles to last level cache (LLC) misses and even help developers understand whether those misses are served by the local or remote NUMA-node.
They fail to give any further insight regarding the remote NUMA-nodes that are on a different board. 

Furthermore, when developing cache-coherent interconnects outside a simulation environment, there are no universal mechanisms available to capture interconnect resource utilization.
Consequently, high-rate capture of fine-grained on-chip interconnect counters would present useful telemetry both to guide interconnect and application development and tuning.
As an example of the first case, on-chip buffering is generally expensive; understanding how much buffering is needed to cover the interconnect round-trip for optimal throughput with appropriate counters, would allow allocation of just enough resources. 
This mismatch occurs since relatively crude activity patterns can be simulated in chip development, and latency isn't modeled at all, with many approximations for simulation speed.
In the latter case, the impact of algorithmic adjustments can be understood at a cache-coherent and high-level ccNUMA viewpoint.

In this paper, we demonstrate the lightweight capture of very high-rate on-chip events (up to \num{2e8} events per second), and visualization in an interactive graphing environment, along with a convenient textual interface for \textit{quick-look} analysis.
We present \numascope, a framework that enables tools that capture events across different cache-coherent interconnects in large ccNUMA systems. 
\numascope is \opensource and based on a modular software architecture, that makes it easily extendable.
\numascope not only enables the user to see at a high-level and quantify suboptimal application (or the system as a whole) behavior, but it also enables identification of \textit{why} it is occurring, from the system's perspective, e.g. because some buffer queues are full.
This information enables the observer to understand whether a high rate of events observed using the tool also happens to saturate some hardware resource, potentially leading to different scheduling decisions or even different hardware configurations.

In detail, in this paper we make the following contributions:
\begin{itemize}
    \item We introduce \numascope, a framework that enables tools that perform real-time and \textit{postmortem} monitoring of cache-coherent interconnects, using a mechanism for high-rate event capture;
    \numascope provides both a graphical and a textual interface that visualize the captured events.
    \item We present how using \numascope, we were able to detect a performance issue in the GNU C Compiler 4.8.5 OpenMP implementation and compare its performance before and after applying a fix.
    \item We evaluate \numascope, on a 6-board ccNUMA system, using the NAS Parallel Benchmarks (NPB)~\cite{nas} EP benchmark, reporting its overhead for various sample rates, starting from a high sample rate at 1000Hz and going down to 1Hz; our evaluation shows that the overhead ranges from 0.27\% at 1Hz sampling, to 5\% at 1000Hz sampling.
\end{itemize}


\section{Related Work}

\subsection{Hardware Counters Access}
Various tools and libraries that enable gathering metrics from hardware counters are readily available~\cite{perf, papi, likwid}.
However these tools and libraries are typically able to gather metrics only within a single board, failing to support large ccNUMA systems.
Furthermore, to get access to NUMA-related metrics, e.g. number of remote accesses, the tools rely on processor-specific events that users need to pass as parameters to the tool.
Identifying and understanding the events that are of interest is not trivial; it requires careful studying of the architecture's manual and understanding of the architecture itself~\cite[\S 3]{memphis}.
To make matters worse, stable OS distributions (e.g. CentOS) usually ship with older kernel versions.
As a result, \texttt{perf}, the most popular profiling tool shipped with Linux, usually lacks support for the performance counters added to newer processors.
For Intel processors \texttt{PMU-tools}~\cite{pmu-tools} have been developed to mitigate this issue.
\numascope compared to \texttt{perf} is more easily extendable and does not rely on the latest kernel to work, nor on kernel patches to extend it.
\numascope can also take advantage of the community support for those tools and libraries and use them for gathering metrics.

\subsection{Gathering Metrics About ccNUMA-interconnects}

Prior work has also focused on gathering micro-architectural events from ccNUMA interconnects~\cite{Origin2000prof, flashpoint}.
Both these works focus, each, on a single proprietary hardware interconnect and are closed-source.
\numascope, in contrast to these approaches~\cite{Origin2000prof, flashpoint}, is \opensource and designed to be modular in order to support different hardware interconnects by implementing a simple software interface.

\subsection{System Monitoring Tools}

\numascope being able to monitor large ccNUMA systems is also related to system monitoring tools.
Most system monitoring tools focus on higher-level metrics such as processor, disk, and network usage~\cite{cloud-monitoring}, as well as even higher-level metrics including service availability.
\numascope, providing \textit{lower-level} metrics, can complement such tools aiding developers to understand the root-cause of deficiencies in their system.

\subsection{Profiling Tools}
By using instruction-based and event-based sampling, prior works have also built profiling tools that are able to correlate code segments with NUMA-related events~\cite{hpctoolkit-numa, memphis, memprof, numap, scaAnalyzer}.
The main goal of these profilers is to help developers detect bottlenecks in their applications, caused by cache hierarchy effects.
However, these tools rely on processor core counters and lack support for large ccNUMA systems.
As a result, profiling applications on large ccNUMA systems gives limited insight into what is happening outside a single board.
Existing NUMA-aware profilers can only tell whether an LLC miss is served by the local NUMA-node or a remote one, with lack of insight regarding accesses to remote NUMA-nodes that are on different boards.
Such information is valuable, since different boards may have different utilization, and a holistic view often aids in optimizing imbalanced workloads.
\numascope can be used in conjunction with such tools to determine the source of excessive memory access time; that would occur when the cycle misses in the local last level cache and the physical address corresponds to a remote NUMA-node.
Without \numascope, core profiling would simply reveal a memory load taking significantly longer.
\vk{I would again say that numascope could be extended somehow if true}

PMU-tools also feature the \texttt{toplev} tool that implements the top-down method~\cite{topdown}.
The top-down method groups performance counters to help developers identify which part of the processor, e.g.\ front-end, back-end, etc.\ is being stressed by their application and find potential bottlenecks.
\numascope and measurements obtained from the ccNUMA interconnects can be used to extend the top-down methodology, and more specifically the \textit{External Memory Bound} group.\\

Overall, \numascope differentiates itself from related work by:
\textit{a)} capturing more diverse events and data,
\textit{b)} being easily extendable, enabling the profiling of different platforms,
\textit{c)} being a framework that enables the creation of specialized monitoring tools on a per case base, and
\textit{d)} being open-source and readily available to the community.
Furthermore, \numascope can be used to complement existing NUMA-aware profiling tools as well as system monitoring tools to aid developers trace the root cause of potential deficiencies in their applications and deployments by exposing low-level metrics obtained from ccNUMA-interconnects.

\section{Background}
 
In this Section we shortly discuss large ccNUMA systems, the importance and the role of ccNUMA interconnects, how they affect performance, and the challenges in profiling applications on systems using them.

\subsection{Large ccNUMA Systems}
\label{sec:background:ccnuma}

\begin{figure}
    \includegraphics[width=\linewidth]{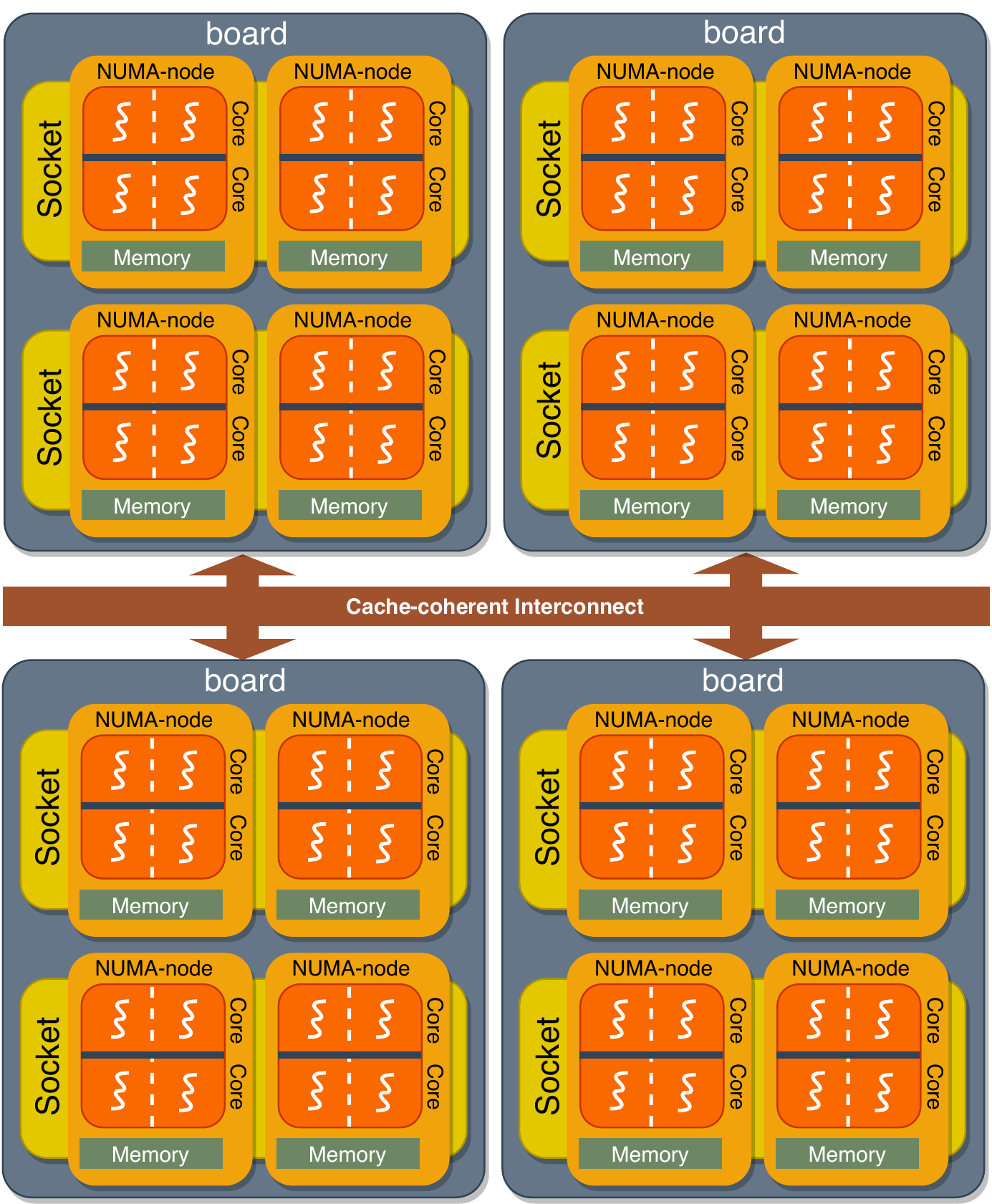}
    \caption{An illustration of a ccNUMA system}
    \label{fig:ccnuma-system}
\end{figure}

Figure~\ref{fig:ccnuma-system} illustrates the layout of a large ccNUMA system which consists of a number of boards or discrete servers; sometimes also called \textit{nodes} (4 in the illustration).
In general, by \textit{large} ccNUMA systems we refer to systems larger than what the processor natively supports without a cache-coherent NUMA interconnect (see Section~\ref{sec:background:interconnects}).
\djb{We sometimes use large ccNUMA, so should decide to use it everywhere or not}.
Each board typically comprises 2--4 processor sockets (2 in the illustration) and a number of NUMA-nodes (4 in the illustration).
Each socket comprises a number of cores and hardware threads (4 and 8 respectively in the illustration).
Typically each hardware thread has it's own level 1 cache and might share higher level caches with other hardware threads on the same chip.
As we move to higher levels in the cache hierarchy, the communication means changes along with the access times (see Figure~\ref{fig:numa-costs}).
Additionally, each processing unit in a NUMA-node has fast access to the local memory of its NUMA-node, while it can still access the rest of the system's memory with varying latency depending on the remote memory location (see Figure~\ref{fig:numa-costs}).

\subsection{Cache-coherent NUMA interconnects}
\label{sec:background:interconnects}
For processors to be optimized for maximum efficiency, they need a common upper limit on the number of sockets they support. This dictates how many bits are allocated to each entry in a \textit{cache directory}, to track location and state of cached lines. This is typically 8 NUMA-nodes (3 bits per cache line)~\cite{processor-design,bkdg}. In addition, in some AMD processors an optimization called HT Assist reserves a fixed amount of the level 3 cache as a \textit{cache directory}, which tracks which NUMA-node holds what cache line in what state~\cite{conway2009blade}. In the AMD Opteron 6000 series the reserved, by HT Assist, space amounts to 2MB out of the 8MB of level 3 cache. Intel processors have a similar amount of level 3 cache partitioned when in multiprocessor mode.

When there is access to a cache line not tracked in the processor's cache directory, an eviction may have to occur to first invalidate a selected cache line to make space for a new entry.
Since it is not known which cache holds state for the newly-tracked cache line, a \textit{probe broadcast} must be sent to all NUMA-nodes.
As the number of NUMA-nodes in a system increases, this broadcast incurs an increasing overhead, reducing application scaling~\cite{snoopbroadcast}.

In large ccNUMA systems that span multiple boards, cache-coherent NUMA interconnects are used to deliver the guarantees of coherency across the whole system.
ccNUMA interconnects overcome the above limitations by being optimized for a considerably larger working set.
This is achieved by tracking many more cache lines and acting as a filter, able to respond to the processor with \textit{directed} probe responses.
This improves scalability, and allows interconnecting multiple smaller (e.g. 2--4 NUMA-node) topologies to offer more processing.

ccNUMA interconnects are typically integrated onto the motherboard, or off-board in a midplane that connects to the processor fabric.
This results in higher access times, compared to intra-board access times, yet significantly lower than accessing storage.
In addition to the cost of visiting a remote memory due to the longer distance and the slower links in the network, the number of cocurrent remote accesses that is supported by the interconnect can also constrain the performance of the system.
If the outstanding remote accesses are less than the processor can generate, applications running on ccNUMA systems may observe increased cache-coherency imposed overheads if not carefully tuned.


\subsection{On-chip counters}
\label{sec:background:counters}

On-chip counters are provided by chip designers to aid software developers to debug and optimize their applications.
These counters offer measurements regarding the events occurring on different hardware modules, such as caches, branch predictors and hardware prefetchers.
For instance, on-chip counters can measure the number of last-level cache misses which indicates the number of memory accesses that end up going off-chip and thus result in higher latency.
Similarly, ccNUMA interconnects offer a variety of on-chip counters that count events specific to the interconnect, such as the number of outstanding remote accesses, the interconnect traffic, etc.
This provides useful feedback to application developers that execute their application at larger scales than previously accessible, for example compared to a dual-socket development system.
On-chip counters may also drive configuration changes in the OS, BIOS and/or firmware in the interconnect after the chip is finalized.
Additionally, they provide crucial feedback to the interconnect developers on how on-chip resources should be allocated in future products.

\subsection{Challenges}

As discussed in Sections~\ref{sec:background:ccnuma} and~\ref{sec:background:interconnects}, minimizing remote accesses is important for improving performance.
Additionally, as discussed in Section~\ref{sec:background:interconnects} it is also important to understand resource utilization in ccNUMA interconnects so as to understand if it presents a bottleneck to the workload and how this could be reduced.

The most common mechanism of accessing on-chip counters is to access registers, directly or indirectly mapped into the application's address space. Access cycles are generated against this mapping which decode to one of the interconnect chips.
Since there are many interconnect chips within the same logical system, the physical address space must have space reserved for access to all chips.
Since the cache-coherent interconnects in large ccNUMA systems are proprietary, there are no standard register layouts, data formats, or even list of counters.
Any tool that implements such accesses is driven by underlying implementation and design-specific details.
As such, it is up to each vendor to develop an interface, documentation, or write a specific tool to access these internal counters.
Finally, it is worth noting that outside the processor cores, physical addresses are always used, preventing understanding where DRAM accesses relate to.

In summary the challenges in profiling applications on large ccNUMA systems are:
\textit{a)} getting access to all the on-chip counters, both from each board and the interconnects;
\textit{b)} interpreting and normalizing the samples per unit time and clock frequency;
\textit{c)} doing the above without imposing significant overhead;
\textit{d)} exposing a way of capturing the data; and
\textit{e)} allowing useful mining and interaction with the collected data, e.g. visually.
%
Our work tackles these challenges and provides a framework that eases the creation of tools able to monitor large ccNUMA systems.

\section{\numascope}

\begin{figure*}
    \includegraphics[clip,width=\linewidth]{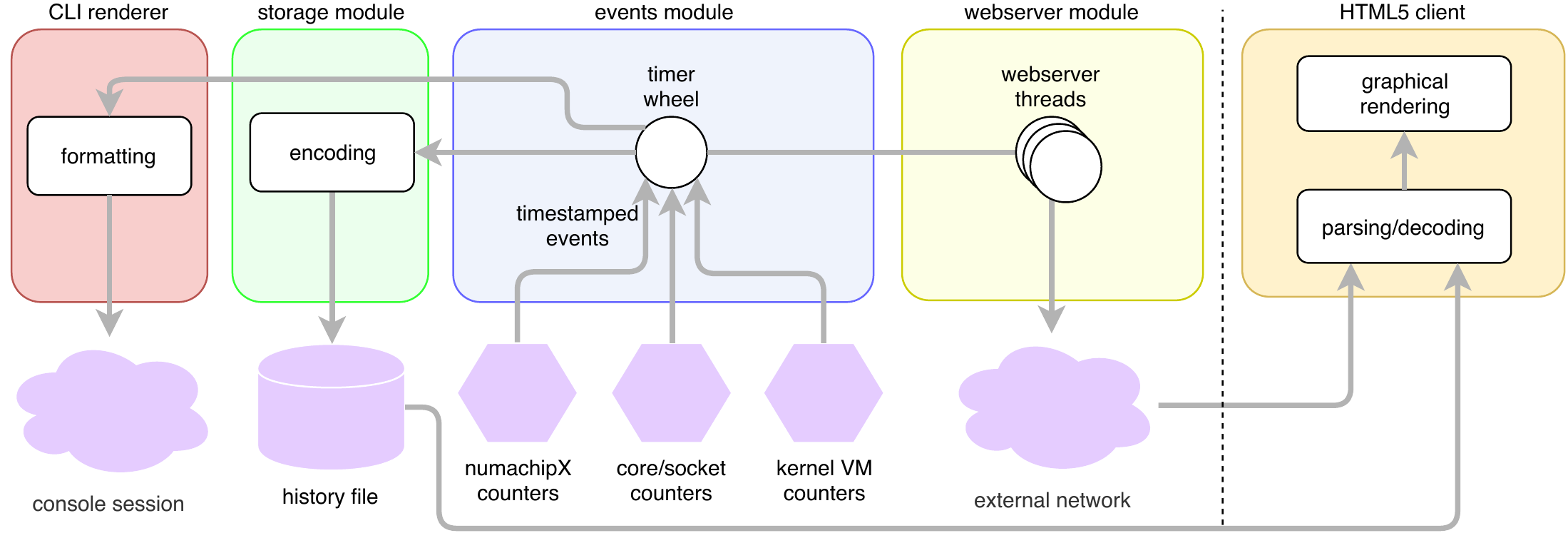}
    \caption{High-level architecture of \numascope}
    \label{fig:architecture}
\end{figure*}

\numascope is a framework that enables the creation of tools that gather and visualize hardware events from different modules on large ccNUMA systems.
To enable extensibility, \numascope has been designed in a modular way, decoupling operating modes, subsystems and capturing of hardware interfaces' metrics, though doesn't need any kernel support.
Figure~\ref{fig:architecture} illustrates the software architecture at a high-level.
As shown, \numascope comprises five modules.

The core module is called the \textit{events module}, responsible for capturing hardware and software events through sampling.
The rest of the modules are optional depending on the usage.
In a typical session, users would first decide if they want to record the data for later analysis or not.
If so, the \texttt{record} mode would be used.
Inspired by the usefulness of the UNIX \texttt{vmstat} command, \numascope offers the \texttt{stat} mode which outputs the selected events to the terminal, allowing the users to readily observe selected metrics.
Finally, the \texttt{live} mode can be used to start a web server to allow interactive graphing of the data in real-time.
%
%
The \textit{storage module} is responsible for storing the captured events when using the \texttt{record} mode.
The \textit{CLI renderer module} is responsible for formatting the captured events on the command line when using the \texttt{stat} mode.
The \textit{web server module} is responsible for interacting with the \textit{HTML5 client} with data in real-time when using the \texttt{live} mode.
Finally, the \textit{HTML5 client} front-end is responsible for providing an interactive graphical interface to visualize in real time the data transmitted by the \textit{webserver module}, or to visualize the data previously captured in \texttt{record} mode.
In the rest of this section, we further discuss each of these modules in more detail.

\subsection{Events module: High-rate event capture}
\label{sec:numascope:events}

By design, in \numascope, event capture is a \textit{hot path}, wherein all state has been setup to allow reading event counters.
\numascope can be extended to support a variety of events, either from software or hardware.
Out of the box, \numascope is capable of capturing kernel virtual-memory events, which are not exposed via standard UNIX tools, such as \texttt{iostat}, \texttt{sar}, \texttt{vmstat} and related, or Linux-specific ones such as \texttt{perf}.
These events aid in the identification of additional overhead when the executing process needs to conduct work in the kernel to take some action; for example, when accessing a page not yet faulted in from a file, or reclaiming pages to satisfy an allocation.
The kernel outputs the event counters in a structured way when reading \texttt{/proc/vmstat}.
\numascope parses the structured text, using one \textit{lseek} and one \textit{read} systemcall, and stores names and values in a dynamic hashtable to allow for \(O(1)\) lookup.
When running in \texttt{live} mode with one HTTP client, \numascope's memory footprint is about 14MiB of pages, 5MiB of which is shared libraries.

\subsubsection{Data Biasing and Interference}
To avoid interfering with applications, \numascope pins its threads to the first NUMA-node.
\numascope uses a thread-pool to handle asynchronous events.
When running in \texttt{live} mode with one HTTP client, \numascope uses 20 threads irrespective of what events are enabled or modules used.
To avoid data-races, mutexes are used to ensure one thread doesn't request access while another thread is changing state of the events.
Additionally, to avoid affecting the counters relating to coherent accesses, all accesses performed by \numascope to remote interconnects for counter collection are only non-coherent register access; this therefore doesn't evict data in the cache hierarchy.
The rate of non-coherent requests is therefore in the order of thousands per second. The volume of remote non-coherent accesses running can be quantified by examining the metric \textit{packets with less than a full cache line sent}; that is because IO accesses to registers are always 32-bit.
During benchmarking, \numascope records a rate of \num{1.90e5} per second as compared to millions of cache-coherent cycles per second, therefore the bias is insignificant and approaching the noise threshold.

\begin{lstlisting}[float,caption=\numascope's interface for adding hardware support,label=lst:interface]
type Sensor interface {
   // human-readable name of hardware
   Name() string
   // checks if hardware is present
   Present() bool
   // maximum sample value for percentages
   Rate() uint
   // number of hardware elements detected
   Sources() uint
   // supported events
   Events() []Event
   // activated enabled events
   Enable(discrete bool)
   // gets names of enabled events
   Headings(mnemonic bool) []string
   // returns samples
   Sample() []int64
   // used to prevent hardware access races
   Lock()
   Unlock()
}
\end{lstlisting}

\subsubsection{The \texttt{Sensor} Interface}
Regarding accessing hardware counters provided by chip developers, \numascope relies on the \texttt{Sensor} interface listed in Listing~\ref{lst:interface} to get access to them.
Hardware counters may be platform specific, so there is no way to make \numascope able to access the counters of all potential hardware platforms without some form of extension.
Since the design of \numascope is intentionally modular, adding support for new hardware counters is achieved by adding a new sub-module to \numascope's events module that implements the given interface for each target hardware platform.

\texttt{Name()} simply returns a human-readable description of the sensor, used in the user interface.
\texttt{Present()} is used to check whether the \texttt{Sensor} is supported on the platform \numascope is running on.
If present, it also gets set up, which typically involves the memory-mapping of the corresponding hardware registers into the process address space.
\texttt{Rate()} returns the maximum number of events per second, used to normalize counter rates to percentages.
\texttt{Sources()} returns the number of hardware elements detected and set up by \texttt{Present()}.
\texttt{Events()} returns an array of \texttt{Event}s supported by the \texttt{Sensor}, for example cycles during the sampling window and number of cache lines of data moved.
\texttt{Enable()} updates internal state to activate the events to be sampled.
The boolean parameter passed to it indicates whether \numascope should sample events per-board or sum over all boards, to increase the brevity of the CLI renderer, or the amount of data transmitted over websockets.
\texttt{Headings()} returns either short event mnemonics or full descriptions useful to the developer, based on the value of the boolean parameter passed to it.
Full descriptions aim to help developers in selecting which events to capture.
\texttt{Sample()} reads out the enabled \texttt{Sensor} events, subtracting from the state held by the previous call to get a rate over time, and returns a slice of the enabled event rates.
Signed 64-bit integers are used, since certain events can legitimately have a negative rate. 
For example kernel virtual memory \textit{swapped-pages} count will increase with VM pressure, giving a positive rate, and decrease when swapped pages are read back in, giving a negative rate.

\subsubsection{Controlling \numascope Dynamically}
The events module, can also be programmatically controlled by writing commands into a UNIX FIFO at \texttt{/run/numascope-ctl}.
This enables controlling \numascope with ease from both shell scripts and programs themselves for finer-grained profiling.
The commands currently supported by \numascope are:
\begin{enumerate}
    \item \texttt{record \textless filename\textgreater} closes any existing recording file and starts recording to given filename,
    \item \texttt{label \textless string\textgreater} inserts a label onto the graph or textual output to mark a notable event for later analysis,
    \item \texttt{pause} suspends recording events without suspending capturing,
    \item \texttt{resume} continues recording events, and
    \item \texttt{interval \textless number\textgreater ms} adjusts the sampling rate of sensors in milliseconds.
\end{enumerate}
\vk{seems that numascope captures system-wide metrics, not per process (as with perf) -- should mention that?}
\fz{Maybe we could add a section "Limitations" and mention that \numascope at its current state cannot capture per process metrics, not attribute such metrics to code segments}
\djb{I think it's prudent we mention that, outside the processor core, all access is to physical addresses, therefore agnostic to the process}.

\subsection{Storage module: Storing data for offline analysis}
Since data may be stored and used even years after gathering them~\cite{isca15}, using a simple and flexible but standards-compliant data storage format has very strong rationale.
As such, both CSV and JSON serve the purpose well, since they are mature and well-supported by many frameworks and tools.
Furthermore, both formats are not dependent on additional information such as a schema as may be expected with XML. 
JSON has three major advantages over CSV though.
Firstly, it can encode structured data, which may be advantageous if the framework is expanded in the future.
Secondly, it is native to JavaScript, so has highly optimized support without requiring external JavaScript libraries.
Finally, this allows it to be consumed directly by graphing frameworks.
In order to avoid making disk IO a bottleneck, the storage module writes samples in a buffer which is periodically flushed to disk by the OS; on Linux, this is every 30 seconds. Sampling at 10Hz, 28KiB of JSON test are stored per second.

Due to the modular approach taken, additional storage modules may be developed to send the data to other storage mechanisms or databases, such as InfluxDB, TimescaleDB, or Apache Cassandra.

\subsection{CLI renderer: Command-line data presentation}

The \texttt{stat} mode of \numascope outputs the selected events to the terminal, allowing the users to readily observe selected metrics, or pipe the output and process it on-the-fly.
The CLI also enables easy monitoring of remote systems without any external dependencies.
Listing~\ref{lst:clishot} shows a snapshot of \numascope's CLI in action.
In this snapshot, we see three interconnect events being monitored; \texttt{n2DirPrbRecv}, \texttt{n2CachelinesSent}, and \texttt{n2CacheRolloutRmpe}.
Each line prints the values obtained by a single sample.
Over time new lines are printed with the latest samples.
In this example, we observe how many \textit{cacheline probes} (snoops) and cache lines are sent over the interconnect per second, and how many level-4 (internal to the interconnect) cache line evictions or \textit{rollouts} occurred per second; this it useful to see how much remote activity the workload is generating to understand how the \textit{working set} interacts with the level-4 cache

\begin{lstlisting}[float,caption=Snapshot of \numascope's CLI,label=lst:clishot]
$ numascope stat
n2DirPrbRecv n2CachelinesSent n2CacheRolloutRmpe ...
          41             7357               6689 ...
         103            21357              19092 ...
       11560           115101              78575 ...
      258872           944292              72099 ...
      387750           598625               6843 ...
      375612           583840               8220 ...
\end{lstlisting}


\subsection{web server module: Live data streaming for interactive graphing}

\numascope provides a standalone graphical interface, directly served via the built-in HTTP web server to allow immediate live updates to be observed.
Additionally to HTTP webserving, data is exposed to clients via websockets, on top of which runs a simple yet custom protocol.
This way, two-way interaction between clients and the web server is possible, overcoming the one way limitation with HTTP.
One example where this is useful, is given multiple live sessions capturing data from \numascope, to control options that are global to all clients. This is chiefly the rate at which \numascope latches the on-chip counters and reads them out. Since these can't be interpolated, all clients must use the same sample interval.
Websockets allow the client to send data back to the web server, which in turn informs all clients of a configuration change.
Since events can be captured at a higher rate than the network round-trip would easily accommodate when in live mode, events are batched and sent down all client websockets at 400 millisecond intervals; this is a good trade-off of network and processing overhead, particularly as the client which needs to re-render graph traces frequently.

\subsection{HTML5 client: Interactive graphing of data}
For the interactive visualization of the measurements, \numascope offers a graphical user interface (GUI) realized through web technologies.
Given the ubiquity of web technologies, how standardized and optimized they are today, developing a native client application would present comparatively an intractable engineering and support burden.
%
%
%
Since the \textit{rate} of events is graphed, the \textit{cycles} counter is used to measure the time period that sampling was running; the values are divided by the time period to get an accurate rate over the unit of time. 
This data is transferred either via websocket or JSON file, and optionally reducd over all the boards to conserve bandwidth or space.
Finally, it is passed to the graphing framework, which renders the graph traces.

\begin{figure*}
    \includegraphics[width=\linewidth]{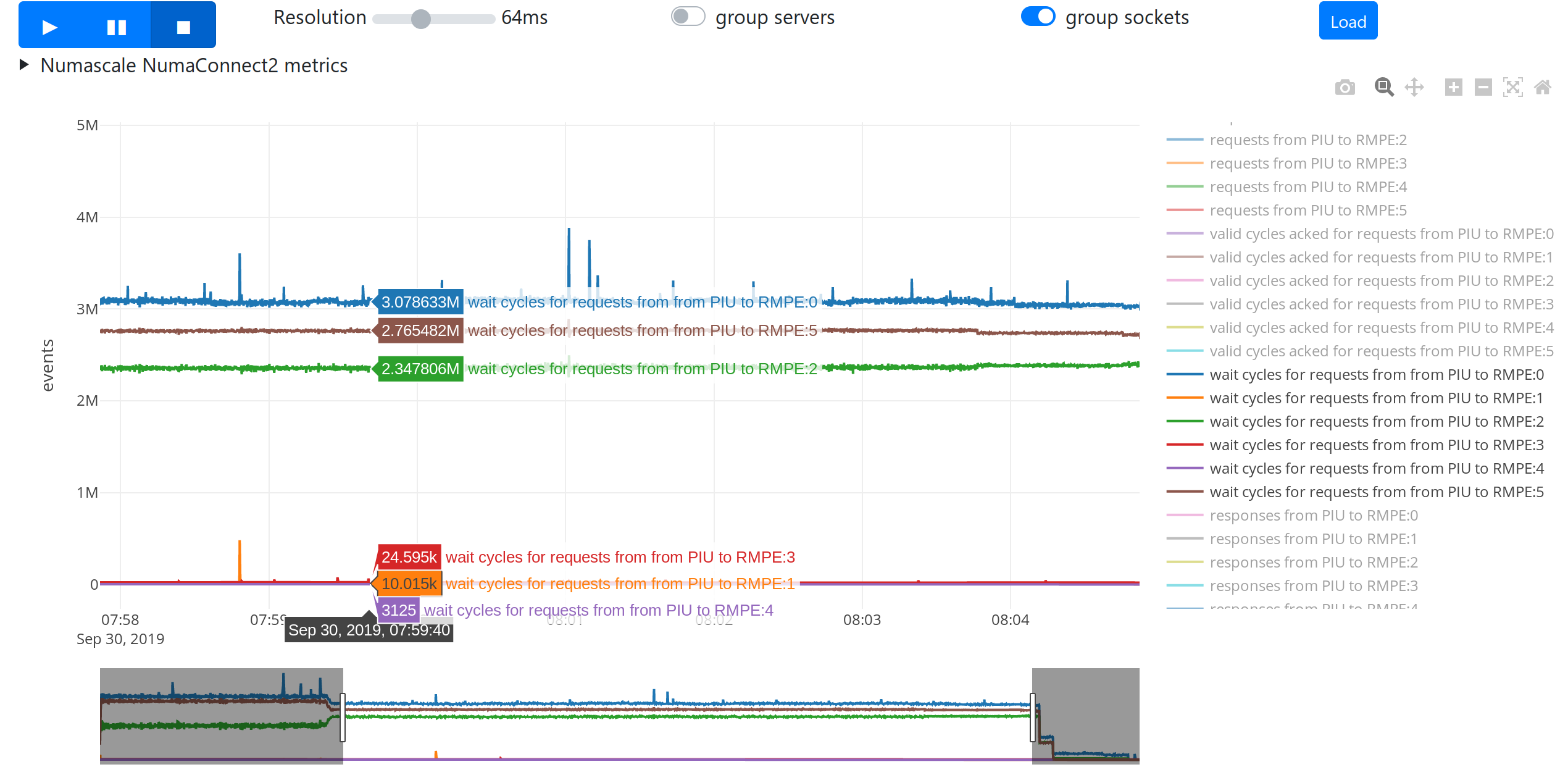}
    \caption{Screenshot of \numascope's GUI}
    \label{fig:guishot}
\end{figure*}

Figure~\ref{fig:guishot} shows a screenshot from \numascope's GUI in action.
At the top of the interface there is a panel with controls allowing the user to adjust the visualization of the data.
The user may control the recording of the data, the sampling rate, and the visualization itself.
Additionally, there is a load button for loading data from a file instead of the web server.
Directly below the control pane, there is an interactive plot visualizing the gathered events in a time-series.
The typical representation of such time-series data is by using one or more line graphs.
The interactive plot allows the user to select which events to be visualized, to focus on a specific time window, and even see the exact value of an event by hovering over the corresponding line on the plot.
Finally, the user can toggle the visibility of traces in the graph by clicking them in the legend.

\section{Evaluation}

To evaluate \numascope, we deploy it on a 6-board, 18-socket AMD Opteron system with 6 ccNUMA interconnects
and use the \textit{Embarrassingly Parallel} (EP) benchmark from the NASA Parallel Benchmark (NPB) suite~\cite{nas} to:
\begin{enumerate}
    \item Demonstrate how \numascope helped us find a performance deficiency in GNU C compiler's (GCC) OpenMP implementation.
    \item Assess the overhead imposed by \numascope.
\end{enumerate}

\begin{figure}
    \centering
    \subfloat[NUMA, one board]{%
        \includegraphics[width=0.47\columnwidth]{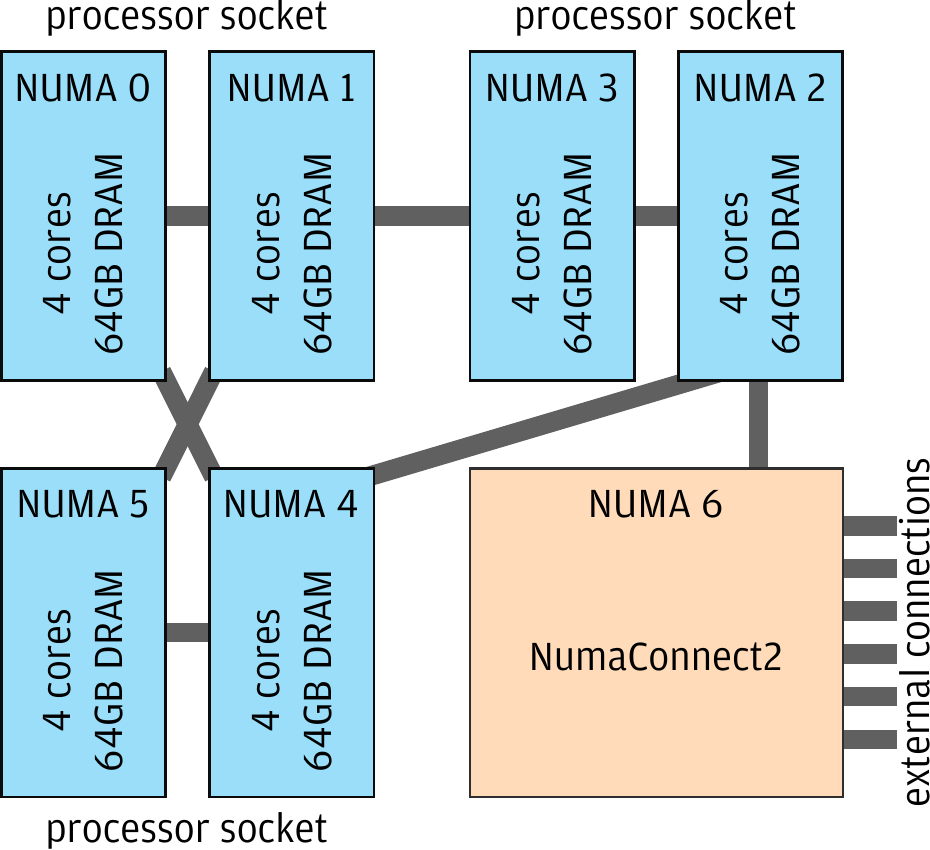}%
        \label{fig:topology-a}%
    }
    \hfil
    \subfloat[Interconnect, all boards]{%
        \includegraphics[width=0.47\columnwidth]{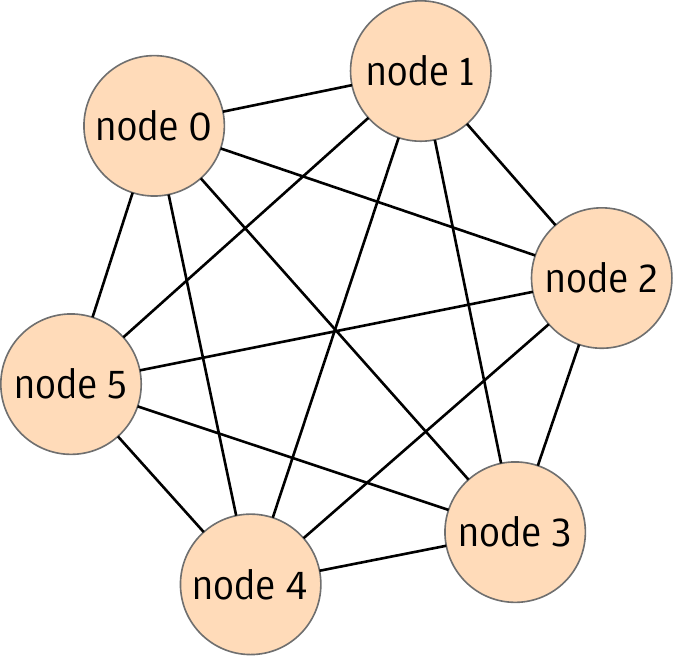}%
        \label{fig:topology-b}%
    }
    \caption{Topology at different levels}
\end{figure}

Each board on the system comprises three AMD Opteron 6328 processors, each featuring 8 cores.
The NUMA topology is shown in Figure~\ref{fig:topology-a}.
The boards are connected using Numascale's NumaConnect2 cache-coherent interconnect, in a point-to-point configuration for single-hop latency, as shown in Figure~\ref{fig:topology-b}.
The links between the NUMA-nodes in the processors supports  6.4GT/s (Giga Transfers) peak throughput; the links between the processors and interconnect supports 3.2GT/s.
Table~\ref{tab:configuration} summarizes the hardware and software configuration of the evaluation platform.

\begin{table}
\normalsize{
    \centering
    \caption{Hardware and Software Configuration}
    \begin{tabular}{r|l}
        CPUs & 18 (3 per board) AMD Opteron 6328 \\
        NUMA-nodes & 36 (6\(\times\)6 per board) \\
        Memory & 2304GiB (384GiB\(\times\)6 per board) \\
        Boards & 6 \\
        CC-interconnect & NumaConnect2 \\
        OS & CentOS 7.6 \\
        Linux Kernel & 4.14.146-NUMASCALE~\cite{kernels} \\
        GCC & 4.8.5 (w/ and w/o patch) \\
    \end{tabular}
    \label{tab:configuration}
}
\end{table}

In our benchmarking, we use \textit{thread pinning} to minimize non-determinism so results are stable and repeatable, and to prevent the kernel scheduler from waking threads up in different areas of the cache hierarchy.
Part of the reason the latter occurs, is from the increased clock jitter due to different boards in the larger ccNUMA system that are driven by different on-board clocks.
One strategy to partially mitigate this is to boot Linux with the \texttt{relax\_domain} parameter set to 3~\cite{os-tips}; this restricts the kernel's process scheduler to limit the search for an unused core to a particular level in the hierarchy (a super-set of the NUMA hierarchy).
To pin the application threads linearly to the core numbers with OpenMP, the environment variable \texttt{OMP\_PIN\_THREADS} is set to \texttt{TRUE}.
Since the evaluation is performed on a dedicated system, in conjunction with thread pinning, the only measurement noise comes from OS housekeeping activities.

\subsection{Numascale NumaConnect2 support}

The Numascale NumaConnect2 ccNUMA interconnect allows interconnecting multiple AMD Opteron 6300 processor boards, providing a level-4 cache and full directory to reduce global traffic.
It allows point-to-point, 3D torus or arbitrary fabric topology which it load-balances over, using open-source firmware~\cite{firmware}.
NumaConnect2 lists all the types of cycles entering and leaving the card, and has counters at most interfaces between functional blocks to accrue how many clock cycles were consumed by waiting.
This includes cache line probes, cache line transfers, IO transfers, as well as level-4 cache hits, misses, evictions and time spent waiting for resources internally in the interconnect.
To expose to \numascope the hardware events gathered by NumaConnect2 we implement the \texttt{Sensor} interface described in Section~\ref{sec:numascope:events} as a new submodule.

The NumaConnect2 counters profiled by \numascope fall into two categories.
Firstly, \textit{extrinsic events}, which include the types of traffic going into the NumaConnect2 from both the processor and interconnect fabric sides.
Understanding this traffic may reveal if an application is exploiting or thrashing the cache hierarchy, which is useful in algorithm design and application tuning.
Workloads thrashing the cache hierarchy would result in considerably more access over the interconnect, for example random access to a working set somewhat larger than the caches. 
Applications exploiting the cache hierarchy would generate considerably less activity on the interconnect either by accessing a working set that fits in the caches, or by accessing local data.
Secondly, the \textit{intrinsic events} which includes internal resource usage, or counting cycles for which a request is stalled waiting for resources.
This helps understand if and when bottlenecks occur in the interconnect.






\subsection{Workloads and metrics}
The NASA Parallel Benchmarks~\cite{nas} present a range of numerical problems with varying characteristics. 
Additionally, varying problem sizes can be configured and the number of threads of execution can be arbitrarily specified, so as to make full use of resources.

Using the NPB EP benchmark allows us a means to validate \numascope.
The EP class D benchmark generates \num{2e36} pairs of pseudorandom numbers with a Gaussian distribution in a 1.1GiB distributed memory array.
In this paper, we analyze the amount of time the interconnect is waiting for internal resources.


From a whole-system perspective, it would be intuitive to measure the number of remote outstanding transactions at the interconnect.
Since this is limited in hardware to 16, we can not get insight into the end-to-end queuing occurring from the processor cores to the remote DRAM.
Instead, a linear measure of queuing is ``wait cycles for requests from SIU to RMPE''; this measures the number of clock cycles spent for cache-coherent requests traveling from the NumaConnect2 \textit{Scalable Coherent Interface} (SCI) \textit{Interface Unit} to the \textit{Remote Memory Processing Engine}; this shows exactly how long memory access was stalled because the interconnect was busy.

\begin{figure*}
    \centering
    \subfloat[Without stack fix]{%
        \includegraphics[width=\columnwidth]{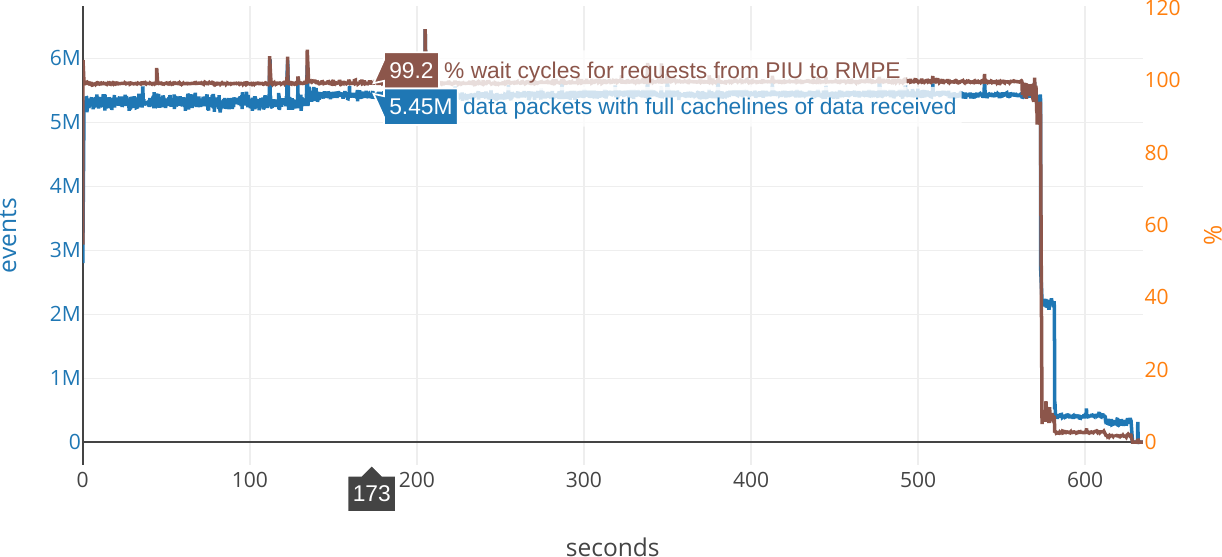}%
        \label{fig:waitcycles-a}%
    }
    \hfil
    \subfloat[With stack fix~\cite{gcc-patch}]{%
        \includegraphics[width=\columnwidth]{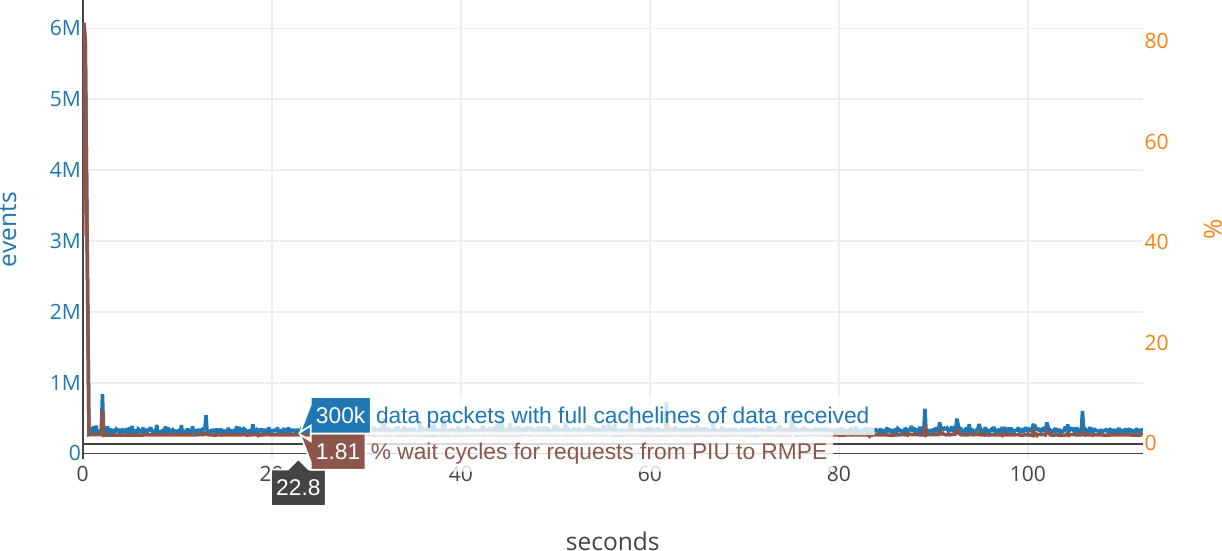}%
        \label{fig:waitcycles-b}%
    }
    \caption{NPB EP class D pinned OpenMP wait cycles}
    \label{fig:waitcycles}
\end{figure*}

\subsection{GCC's OpenMP implementation deficiency}
During execution of the NASA Parallel benchmarks, we observed using \numascope, that the benchmarks were causing an unexpectedly high number of cycles waiting for resources in the interconnect, even with the NPB \textit{Embarrassingly Parallel} (EP) benchmark.
As implied by its name, the EP benchmark has no \textit{data dependencies} among parallel threads, so minimal communication is expected to take place, and memory accesses should be mostly local or cached.
Figure~\ref{fig:waitcycles-a} plots, on the y-axis, the number of cache-lines transferred over the interconnects (blue line), along with the percentage of the ``wait cycles for requests from SIU to RMPE'' over the theoretical maximum of the system (brown line); the x-axis is time in seconds.
Given the 200MHz core clock speed of NumaConnect2 the maximum number of ``wait cycles for requests from SIU to RMPE'' that can be observed is \num{0.2e9} per interconnect.
To calculate the plotted percentage we use the following equation:
\begin{equation}
    100 \times \sum_{n=1}^{\#interconnects} \frac{\mathrm{wait\_cycles}_n}{0.2e9}
\end{equation}
where \(\mathrm{wait\_cycles}_n\) is the ``wait cycles for requests from SIU to RMPE'' measurement obtained from interconnect \(n\).
As shown in Figure~\ref{fig:waitcycles-a}, there is a constant rate of about 99\% (out of the maximum 600\%) of the total number of clock cycles spent waiting for resources across all six boards.
This observation suggests that one of the six interconnects is acting as the bottleneck, at constant saturation.

\begin{figure}
    \centering
        \includegraphics[width=\linewidth]{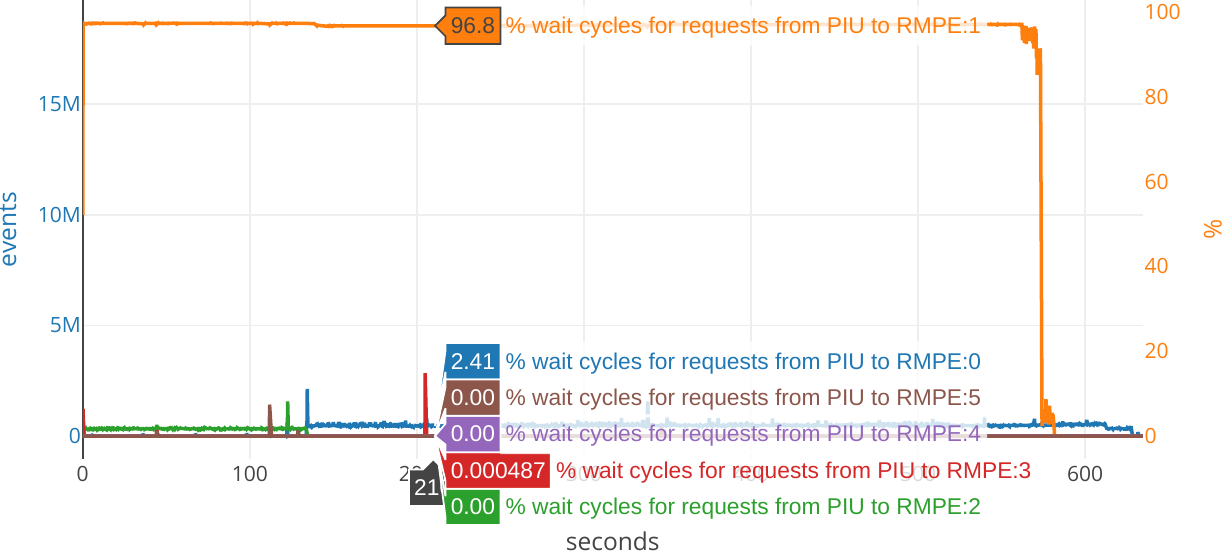}
    \caption{NPB EP class D pinned OpenMP wait cycles per board}
    \label{fig:waitcycles-detail}
\end{figure}

Disabling the grouping of the boards in the user interface, we reveal the wait cycles per board, as shown in Figure~\ref{fig:waitcycles-detail}.
In a well-balanced workload like EP, we would expect similar values for each board.
However, from Figure~\ref{fig:waitcycles-detail}, it can be seen that the wait cycles are almost all from the second board.
The second board sees about 97\% time spend in wait cycles, while the first board sees about 2.5\%, with the other boards experiencing negligible contribution.
This illustrates a \textit{highly imbalanced} access pattern, perhaps one that would fit with a software placement bug.
This ultimately led to the discovery that GNU C compiler's OpenMP thread pinning was working, but was touching each thread's stack in the parent thread.
This resulted in stack pages being page-faulted on the parent-thread's NUMA-node.
Since the stack is intensively used, this generates considerable cross-interconnect traffic.
To mitigate this a workaround was prepared to identify the correct NUMA-node and pin the stack explicitly~\cite{gcc-patch}. 
\vk{the solution is not explained and the reference is not working (on purpose I suppose)}
\fz{Yes the reference is anonymized. The explanation does not really add something to the paper so the reader should normally need to go to the reference and read it. Not sure how to properly handle this. We already describe the solution at high-level.}

Recompiling EP with the patched GCC and rerunning the benchmark, run-time dropped from \(634.4\)s to \(112.1\)s; a factor of \(5.7\times\) speedup.
Correspondingly, the interconnect traffic dropped from a total of 17.8GiB to 1.14GiB; a factor of \(15.6\times\) less interconnect traffic.
Figure~\ref{fig:waitcycles-b} shows the corresponding plot.
We see that in the patched run, number of cache-line transfers reaches \num{6e6} per second during initialization and when the parallel work starts they drop to \num{3.80e5} steady-state.
Correspondingly, the wait time drops to about 2\%.


\begin{figure}
    \centering
    \includegraphics[width=\linewidth]{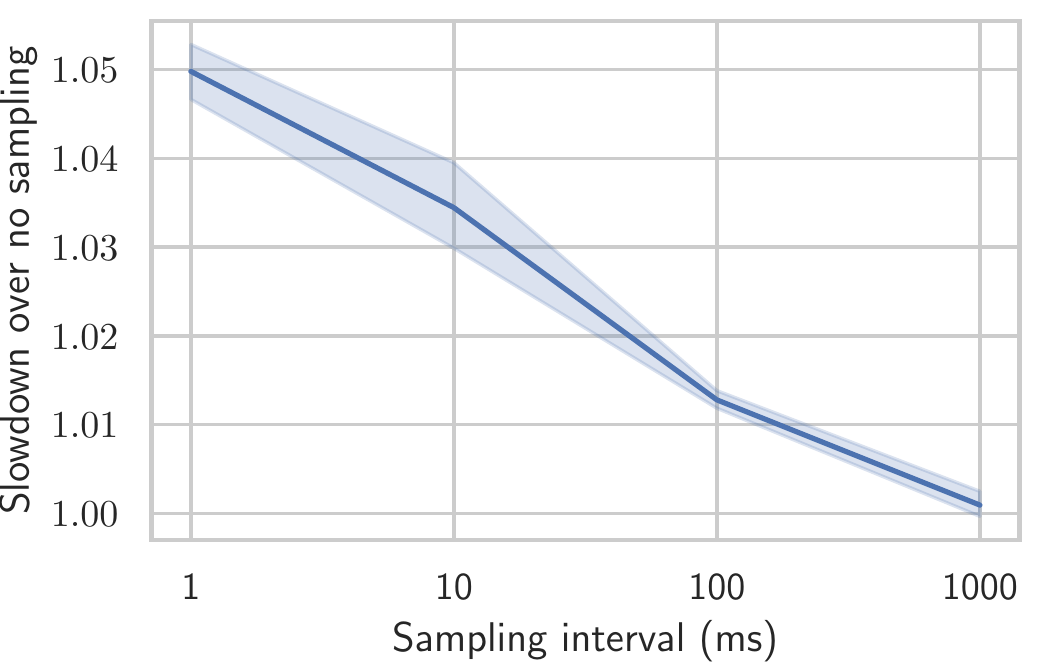}
    \caption{\numascope overhead with various sampling rates}
    \label{fig:overhead}
\end{figure}

\subsection{Tool overhead}
To determine the overhead of \numascope, we execute the NPB EP class D benchmark without using \numascope, and again using \numascope in \texttt{record} mode.
We demonstrate the overhead in \texttt{record} mode since this is the only mode expected to be used for long periods of time, e.g. in production.
\vk{record involves storing time-series? what about storage requirements when running for long period in production?}
\fz{Daniel: can we provide some formula, or rough estimates about the disk-requirements per hour depending on the sampling rate?}
Figure~\ref{fig:overhead} plots the slowdown of the benchmark as we increase the sampling interval from 1ms to 1s.
We observe that the maximum overhead on average is about 5\% with a sample interval of 1ms (1KHz), while the minimum is 0.27\% with a sample interval of 1s (1Hz).
In typical use scenarios, for short period, a sample interval of 100ms gives sufficient resolution for helping to understand application and interconnect behavior, and gives low overhead due to the noise mitigation steps detailed previously, and the access to remote NumaConnect adapters being with non-coherent cycles, therefore \textit{cache oblivious}.
Note that the sampling interval can be dynamically changed the \texttt{/run/numascope-ctl} UNIX FIFO.
That said \numascope can be initially configured to sample at a 100ms interval and if something alerting is observed the sampling interval can be further increased to investigate.


\section{Conclusions}

\numascope is an \opensource~\cite{numascopeGithub} framework that enables the creation of tools that allow the monitoring of large ccNUMA systems.
\numascope is designed to be modular and hardware agnostic, to support a number of ccNUMA interconnects in the market.
By implementing a simple interface, vendors can add support for newer interconnets and expose measurements from hardware counters.
\numascope can then capture various events, from kernel events, to hardware events and visualize them in an interactive GUI.
As demonstrated in this work, tools based on \numascope can help developers understand their applications' bottlenecks and fix them.
\numascope's overhead is shown to be negligible when used with a high sampling interval (1s), and it can go up to about 5\% for a sampling interval of 1ms.
That makes \numascope-based tools production-ready.

\balance
\bibliography{paper}
\bibliographystyle{IEEEtran}
\end{document}